\documentstyle[preprint,aps,epsf]{revtex}

\begin{document}

\draft

%\preprint{REGENSBURG }

\title{$\lowercase{n}$-point functions at finite temperature}

\author{Defu Hou$^{a,b}$, Enke Wang$^{a,}$\cite{address2}, and 
        Ulrich Heinz$^a$} 

\address{
   $^a$Institut f\"ur Theoretische Physik, Universit\"at Regensburg,\\
   D-93040 Regensburg, Germany
}
\address{
   $^b$Institute of Particle Physics, Huazhong Normal University,\\
       Wuhan 430070, P. R. China
}

\date{\today}

\maketitle

\begin{abstract}

We study $n$-point functions at finite temperature in the closed
time path formalism. With the help of two basic column vectors and
their dual partners we derive a compact decomposition of the
time-ordered $n$-point functions with $2^n$ components in terms of
$2^{n-1} -1$ independent retarded/advanced $n$-point functions. This
representation greatly simplifies calculations in the real-time
formalism. 
 
\end{abstract} 


\narrowtext

%%%%%%%%%%%%%%%%%%%%%%%%%%%%%%%%%%%%%%%%%%%%%%%%%%%%%%%%%%%%%%%%%%
\section{Introduction}  
\label{sec1}
%%%%%%%%%%%%%%%%%%%%%%%%%%%%%%%%%%%%%%%%%%%%%%%%%%%%%%%%%%%%%%%%%%

Finite temperature field theory has been extensively studied for a
large number of problems in particle physics, solid state physics and
the physics of the early universe. There exist real-time (RTF) and 
imaginary-time (ITF) formulations of finite temperature field
theory~\cite{kap,rep-145,Chou,Bellac}. The real-time formalism again
comes in several variations~\cite{peterL}. To study the relations
between those different formulations of the theory has been an
interesting subject over the years~\cite{Kob,Kob1,evans,TimE,Gu,Aur}. 
In Ref.~\cite{evans} Evans gave relations between the RTF 3-point
Green functions and their analytically continued ITF counterparts.
In Ref.~\cite{TimE} he noted that the situation was more complicated
for 4-point and higher order $n$-point functions. Taylor studied
the corresponding relations for 4-point functions in~\cite{taylor}.
In \cite{umezawa,PeterH} a column vector calculus for thermo-field
dynamics was presented which simplifies the calculation of Feynman
diagrams for matrix valued propagators. More recently in
Ref.~\cite{u&m} the relation between the time-ordered and 
retarded/advanced \cite{Aur} 3-point functions was reexamined in the
closed time-path (CTP) formalism \cite{Chou,PeterH}, and a simple
decomposition of the 8-component real-time vertex tensor in terms 
of retarded/advanced 3-point functions was derived using outer
products of the 2-component column vectors introduced in
\cite{umezawa,PeterH}. It was shown in \cite{u&m,hh} that, due to
orthogonality relations between the column vectors, this
representation greatly simplifies calculations in the real-time
formalism. 

The purpose of this short report is to establish relations between
the time-ordered and retarded/advanced $n$-point functions by using
the column vector technique. We will generalize the work in
Ref.~\cite{u&m} on the 3-point functions to $n$-point functions. In
doing so we shall show that all $2^n$ time-ordered functions can be 
expressed through linear combinations of $2^{n-1}-1$ independent 
retarded/advanced functions.

In Sec.~\ref{sec2} we will review the column vector representation of
the two-point function and study how it is decomposed into 
retarded/advanced propagators. In Sec.~\ref{sec3} we will study the 
analogous decomposition for the $3$-point functions. 
General $n$-point functions will be studied in Sec.~\ref{sec4}. Some
conclusions are presented in Sec.~\ref{sec5}. 

%%%%%%%%%%%%%%%%%%%%%%%%%%%%%%%%%%%%%%%%%%%%%%%%%%%%%%%%%%%%%%%%%%
\section{2-point functions}
\label{sec2}
%%%%%%%%%%%%%%%%%%%%%%%%%%%%%%%%%%%%%%%%%%%%%%%%%%%%%%%%%%%%%%%%%%

To establish our notation and for later use we first consider the 
single-particle propagator for a bosonic field theory. In real time,
the propagator has $2^2=4$ components since each of the two fields can
take values on either branch of the real-time contour
\cite{rep-145,Chou}. We generally follow the notation in
Ref.~\cite{PeterH}.  

The 4 components of the propagator can be combined into a $2 \times 2$
matrix   
 \begin{equation}
 \label{2x2}
   D = \left(  \matrix {D_{11} & D_{12} \cr
                        D_{21} & D_{22} \cr} \right) \, ,
 \end{equation}
which can be rewritten \cite{umezawa,PeterH} in terms of an outer
product of 2-component column vectors:
 \begin{eqnarray}
   D(p) &=& D_R(p) {1\choose 1}{\otimes}{1+n(p_0)\choose n(p_0)} 
          - D_A(p) {n(p_0)\choose 1+n(p_0)}{\otimes}{1\choose 1}\, .
 \label{Dp}
 \end{eqnarray} 
Here 
 \begin{equation}
 \label{Bose}
   n(p_0) = {1 \over e^{\beta p_0} -1}\, ,\qquad 
   n(-p_0) = - \bigl( 1 + n(p_0) \bigr)
 \end{equation}
is the thermal Bose-Einstein distribution, and $D_{R,A}$ are the
retarded and advanced propagators 
 \begin{equation}
 \label{RA}
   D_R = D_{11}-D_{12}\, \qquad 
   D_A = D_{11}-D_{21}\,.
 \end{equation}
Their spectral representations are
 \begin{equation}
 \label{spectral}
   D_R(p) = \int_{-\infty}^\infty \frac{d\omega}{2\pi}
          {\rho_-(\omega,{\bf p}) \over
           p_0 - \omega + i\epsilon} \, , \qquad 
   D_A(p) = D^*_R(p).
 \end{equation}
$\rho_-(p)$ is the (real) spectral density in terms of which all
propagator components can be expressed via spectral integrals.

Defining the following two column vectors
 \begin{eqnarray}
  \bbox{e}_R(p) = { 1+n(p_0) \choose n(p_0)}\,,  \qquad
  \bbox{e}_A(p) = -{ 1 \choose 1}\,, 
 \label{basis}
 \end{eqnarray}
Eq.~(\ref{Dp}) can be written as
 \begin{eqnarray}
 \label{Dp1}
   D(p) = -\Bigl( D_R(p)\, \bbox{e}_A(-p){\otimes}\bbox{e}_R(p)
        + D_A(p)\, \bbox{e}_R(-p){\otimes}\bbox{e}_A(p)\Bigr)\, .
 \end{eqnarray} 
We also define the ``dual'' vectors
 \begin{equation}
  \tilde{\bbox{e}}_R(p) = { 1 \choose -1}\,,  \qquad
  \tilde{\bbox{e}}_A(p) = { n(p_0) \choose -1-n(p_0)}\,,
 \label{dual}
 \end{equation}
which satisfy ($\tau_3$ is the third Pauli matrix)
 \begin{equation}
  \tilde{\bbox{e}}_{R,A}(p) = -\tau_3\, \bbox{e}_{A,R}(-p)\,.
 \label{dual1}
 \end{equation}
As a rule of thumb the vectors $\bbox{e}_{R,A}$ are associated with 
legs which carry outflowing momenta while the $\tilde{\bbox{e}}_{R,A}$
are associated with inflowing momenta. If the propagator
$D(p)=G^{(2)}(p,-p)$ is represented by a Feynman diagram in which the
momentum $p$ flows from left to right, the left leg carries the
inflowing momentum $p$ (i.e. the outflowing momentum $-p$) while the
right leg carries the inflowing momentum $-p$ (i.e. the outflowing
momentum $p$). This structure is reflected by the signs of the
momentum arguments of the column vectors $\bbox{e}_{R,A}$ in
Eq.~(\ref{Dp1}).

With the column vector contraction rule \cite{umezawa,PeterH}
 \begin{eqnarray}
  { a_1 \choose a_2}\cdot { b_1 \choose b_2} = a_1 b_1 + a_2 b_2
 \label{contraction}
 \end{eqnarray}
the basis vectors (\ref{basis}) and their dual partners (\ref{dual}) 
satisfy the orthogonality relations
 \begin{eqnarray}
   \bbox{e}_\alpha(p)\cdot\tilde{\bbox{e}}_\beta(p) =
   \delta_{\alpha\beta} \qquad (\alpha,\beta=R,A) .
 \label{ortho}
 \end{eqnarray}
Throughout this paper we will use latin letters $a,b,c,\dots =1,2$ for
the usual thermal indices and greek letters
$\alpha,\beta,\gamma,\dots=R,A$ for retarded/advanced indices. 

The self-energy $\Pi$, defined through the Schwinger-Dyson equation 
$D^{-1} = D_0^{-1} -\Pi$, is given by \cite{u&m}
 \begin{eqnarray}
   \Pi(p) &=& \Pi_R(p) {1\choose -1}{\otimes}{1+n(p_0)\choose -n(p_0)} 
          -\Pi_A(p) {n(p_0)\choose -1-n(p_0)}{\otimes}{1\choose -1}\, ,
 \label{Pip}
 \end{eqnarray} 
which can be rewritten as
 \begin{eqnarray}
 \label{Pip1}
   \Pi(p,-p) = -\Bigl( \Pi_R(p)\, \tilde{\bbox{e}}_R
                         (p){\otimes}\tilde{\bbox{e}}_A(-p)
                  + \Pi_A(p)\, \tilde{\bbox{e}}_A
                         (p){\otimes}\tilde{\bbox{e}}_R(-p)\Bigr)\, .
 \end{eqnarray} 
The two momentum arguments denote the two inflowing momenta in the
legs connected to the self-energy. Let us define its $R/A$ components 
via
 \begin{equation}
 \label{defnPi}
   \Pi_{\alpha\beta}(p,-p) \equiv
   \Bigl(\bbox{e}_\alpha(p){\otimes}\bbox{e}_\beta(-p)\Bigr)
   \bullet \Pi(p,-p) \qquad  (\alpha,\beta=R,A)\, .
 \end{equation}
The $\bullet$ denotes the contraction between the 2 vectors
$\bbox{e}$ in the outer product shown in Eq.~(\ref{defnPi})
with the corresponding 2 dual vectors $\tilde{\bbox{e}}$
carrying the same momentum argument in Eq.~(\ref{Pip1}). Using
(\ref{ortho}) one finds 
 \begin{equation}
 \label{PiRA}
   \Pi_{RR}=\Pi_{AA}=0\, , \qquad 
   \Pi_{RA} = \Pi^*_{AR} = -\Pi_R = -\Pi^*_A\, .
 \end{equation}
Note that this differs from the notation used in Ref.~\cite{Aur}
who give $\Pi_{RA}=\Pi_{AR}=0$ and $\Pi_{RR} = \Pi_R = \Pi^*_{AA}$. 
The reason for this discrepancy is that we use the convention that
both momenta in $\Pi(p,-p)$ are inflowing, and our $R$ and $A$ indices
indicate retarded and advanced boundary conditions for these momenta
(see below). Aurenche and Becherrawy, on the other hand, use for the
two-point function (contrary to their convention for the 3-point
function which agrees with ours) one in- and one outgoing
momentum. Following (\ref{dual1}) this implies that on the leg with
the outgoing momentum the $R$ and $A$ indices should be reversed
relative to our notation, in agreement with the above findings. 
 
%%%%%%%%%%%%%%%%%%%%%%%%%%%%%%%%%%%%%%%%%%%%%%%%%%%%%%%%%%%%%%%%%%%%%
\section{3-point vertex functions}
\label{sec3}
%%%%%%%%%%%%%%%%%%%%%%%%%%%%%%%%%%%%%%%%%%%%%%%%%%%%%%%%%%%%%%%%%%%%%

The RTF $n$-point Green functions have $2^n$ components $G_{a_1a_2
  \dots a_n}(p_1,p_2,\dots,p_n)$ where $a_i=1$ or 2. We denote the
truncated $n$-point function, from which the $n$ external propagators
have been removed, by $\Gamma_{a_1a_2 \dots a_n}(p_1,p_2,\dots,p_n)$
and call it the {\em $n$-point vertex}. We adopt the convention that
all momenta flow into the $n$-point vertex such that
$p_1+p_2+\dots+p_n=0$ due to energy-momentum conservation. 

We will show here that the $n$-point vertex functions can be
conveniently expressed as sums of outer products of $n$ column vectors 
in a very similar way as Eq.~(\ref{Pip1}). For the {\em connected}
3-point function such a representation was given in Eq.~(35) of
Ref.~\cite{u&m}. The analogous representation of the {\em truncated} 
3-point vertex is obtained by reversing the signs of all lower
components in the column vectors~\cite{fn1}:
 \begin{eqnarray}
   \Gamma(p_1,p_2,p_3) &=& \Bigl(\Gamma_{abc}(p_1,p_2,p_3)\Bigr) 
 \nonumber\\
   &=& \Gamma_{Ri}(p_1,p_2,p_3) \,
   {1\choose-1}{\otimes}{n_2\choose-1-n_2}{\otimes}{n_3\choose-1-n_3} 
 \nonumber\\
   &-& {(1+n_2)(1+n_3)-n_2n_3 \over (1+n_1) - n_1} 
   \Gamma_{Ri}^*(p_1,p_2,p_3) \,
   {n_1\choose-1-n_1}{\otimes}{1\choose-1}{\otimes}{1\choose-1}
 \nonumber\\ 
   &+& \Gamma_R(p_1,p_2,p_3) \,
   {n_1\choose-1-n_1}{\otimes}{1\choose-1}{\otimes}{n_3\choose-1-n_3} 
 \nonumber\\
   &-& {(1+n_1)(1+n_3)-n_1n_3 \over (1+n_2) - n_2} 
   \Gamma^*_R(p_1,p_2,p_3) \,
   {1\choose-1}{\otimes}{n_2\choose-1-n_2}{\otimes}{1\choose-1}
 \nonumber\\
   &+& \Gamma_{Ro}(p_1,p_2,p_3) \,
   {n_1\choose-1-n_1}{\otimes}{n_2\choose-1-n_2}{\otimes}{1\choose-1} 
 \nonumber\\
   &-& {(1+n_1)(1+n_2)-n_1n_2 \over (1+n_3) - n_3} 
   \Gamma_{Ro}^*(p_1,p_2,p_3) \,
   {1\choose-1}{\otimes}{1\choose-1}{\otimes}{n_3\choose-1-n_3}\, .
 \label{Gamma3}
 \end{eqnarray}
Here $n_i \equiv n(p_i^0),\ i=1,2,3$. Using (\ref{dual}) this can be
rewritten as 
 \begin{eqnarray}
   \Gamma(p_1,p_2,p_3) &=& \Gamma_{RAA}(p_1,p_2,p_3)\,
   \tilde{\bbox{e}}_R(p_1){\otimes}\tilde{\bbox{e}}_A
                     (p_2){\otimes}\tilde{\bbox{e}}_A(p_3)
 \nonumber\\ 
   &+&\Gamma_{ARR}(p_1,p_2,p_3) \,
   \tilde{\bbox{e}}_A(p_1){\otimes}\tilde{\bbox{e}}_R
                     (p_2){\otimes}\tilde{\bbox{e}}_R(p_3)
 \nonumber\\ 
   &+& \Gamma_{ARA}(p_1,p_2,p_3)\,
   \tilde{\bbox{e}}_A(p_1){\otimes}\tilde{\bbox{e}}_R
                     (p_2){\otimes}\tilde{\bbox{e}}_A(p_3)
 \nonumber\\ 
   &+& \Gamma_{RAR}(p_1,p_2,p_3)\,
   \tilde{\bbox{e}}_R(p_1){\otimes}\tilde{\bbox{e}}_A
                     (p_2){\otimes}\tilde{\bbox{e}}_R(p_3)
 \nonumber\\
   &+& \Gamma_{AAR}(p_1,p_2,p_3)\,
   \tilde{\bbox{e}}_A(p_1){\otimes}\tilde{\bbox{e}}_A
                     (p_2){\otimes}\tilde{\bbox{e}}_R(p_3)
 \nonumber\\ 
   &+& \Gamma_{RRA}(p_1,p_2,p_3)\,
   \tilde{\bbox{e}}_R(p_1){\otimes}\tilde{\bbox{e}}_R
                     (p_2){\otimes}\tilde{\bbox{e}}_A(p_3)\, ,
 \label{Gamma3RA}
 \end{eqnarray}
where
 \begin{equation}
 \label{defn}
   \Gamma_{\alpha\beta\gamma}(p_1,p_2,p_3) \equiv
   \Bigl(\bbox{e}_\alpha(p_1){\otimes}\bbox{e}_\beta
                     (p_2){\otimes}\bbox{e}_\gamma(p_3)\Bigr)
   \bullet \Gamma(p_1,p_2,p_3) \qquad
   (\alpha,\beta,\gamma=R,A)\, .
 \end{equation}
The indices $R,A$ of the 3-point functions on the r.h.s. of
(\ref{Gamma3RA}) indicate whether a retarded or advanced propagator
($D_R$ or $D_A$, respectively) is attached to the corresponding leg. 

As shown in Ref.~\cite{Aur}, the retarded/advanced 3-point
functions on the r.h.s. of (\ref{Gamma3RA}) are straightforward
analytical continuations $i\omega_j \to p^0_j+i\epsilon_j\ 
(j=1,2,3)$ of the ITF 3-point function, with $\epsilon_j>0$ 
for each momentum corresponding to a leg labelled by $R$ and 
$\epsilon_j<0$ for each momentum corresponding to a leg labelled by
$A$. Momentum conservation requires $\epsilon_1+\epsilon_2 +
\epsilon_3 = 0$ which implies $\Gamma_{AAA}=\Gamma_{RRR}=0$
\cite{Aur}. This explains why these two $R/A$ functions do not appear
on the r.h.s. of (\ref{Gamma3RA}). While $\Gamma_{AA\dots A}=0$
follows directly from the general identity \cite{Chou}
 \begin{equation}
 \label{sum}
   \sum_{a_1,a_2,\dots,a_n=1}^2 \Gamma_{a_1a_2\dots a_n} = 0\, ,
 \end{equation}
the vani\-shing of the $n$-point vertex with only retarded legs,
$\Gamma_{RR\dots R}=0$, is a result \cite{WH98} of the KMS condition
and is therefore only true in global thermal equilibrium. 

Comparison of (\ref{Gamma3RA}) with (\ref{Gamma3}) yields the
identifications
 \begin{eqnarray}
 \label{comp}
   \Gamma_{Ri} = \Gamma_{RAA}, \quad
   \Gamma_R = \Gamma_{ARA}, \quad
   \Gamma_{Ro} = \Gamma_{AAR}
 \end{eqnarray}
as well as the identities
 \begin{mathletters}
 \label{KMS1}
 \begin{eqnarray}
 \label{KMS1a}
   \Gamma_{ARR}(p_1,p_2,p_3) &=& 
   -{(1+n_2)(1+n_3)-n_2n_3 \over (1+n_1) - n_1}
   \Gamma^*_{RAA}(p_1,p_2,p_3)\, ,
 \\
 \label{KMS1b}
   \Gamma_{RAR}(p_1,p_2,p_3) &=& 
   -{(1+n_1)(1+n_3)-n_1n_3 \over (1+n_2) - n_2}
   \Gamma^*_{ARA}(p_1,p_2,p_3)\, ,
 \\
 \label{KMS1c}
   \Gamma_{RRA}(p_1,p_2,p_3) &=& 
   -{(1+n_1)(1+n_2)-n_1n_2 \over (1+n_3) - n_3}
   \Gamma^*_{AAR}(p_1,p_2,p_3)\, .
 \end{eqnarray}
 \end{mathletters}
The latter are manifestations of the general relation \cite{WH98} 
 \begin{eqnarray}
 \label{KMSgen}
  \Gamma_{\alpha_1\dots\alpha_n}(p_1,\dots,p_n) &=& (-1)^{n+1}
  {\prod\limits_{\{i\vert\alpha_i=R\}} (1+n_i) \over
              \prod\limits_{\{i\vert\alpha_i=A\}} n_i}\,  
  \Gamma^*_{\bar\alpha_1\dots\bar\alpha_n}(p_1,\dots,p_n) 
 \nonumber\\
  &=& (-1)^n {\prod\limits_{\{i\vert\alpha_i=R\}} (1+n_i) 
             -\prod\limits_{\{i\vert\alpha_i=R\}} n_i
              \over
              \prod\limits_{\{i\vert\alpha_i=A\}} (1+n_i) 
             -\prod\limits_{\{i\vert\alpha_i=A\}} n_i}\,  
  \Gamma^*_{\bar\alpha_1\dots\bar\alpha_n}(p_1,\dots,p_n) 
 \end{eqnarray}
which results from the KMS condition in thermal equilibrium. Here
$\bar\alpha=A,R$ if $\alpha=R,A$, respectively. The second equality in
(\ref{KMSgen}) is easily proven using the identity (\ref{A3}).
 
%%%%%%%%%%%%%%%%%%%%%%%%%%%%%%%%%%%%%%%%%%%%%%%%%%%%%%%%%%%%%%%
\section{$\lowercase{n}$-point vertex functions for $\lowercase{n}\geq 4$}
\label{sec4}
%%%%%%%%%%%%%%%%%%%%%%%%%%%%%%%%%%%%%%%%%%%%%%%%%%%%%%%%%%%%%%%

Running through the manipulations of the previous Section in reverse
order it is now easy to generalize Eq.~(\ref{Gamma3}) to higher order
$n$-point functions. We illustrate the procedure for $n$=4;
its generalization to arbitrary $n$ is then straightforward.

Using $\Gamma_{AAAA}(p_1,p_2,p_3,p_4) = 0 =
\Gamma_{RRRR}(p_1,p_2,p_3,p_4)$, the analogue of (\ref{Gamma3RA})
reads
 \begin{eqnarray}
   \Gamma(p_1,p_2,p_3,p_4) &=& 
   \Gamma_{RAAA}(p_1,p_2,p_3,p_4)\,
   \tilde{\bbox{e}}_R(p_1){\otimes}\tilde{\bbox{e}}_A
                     (p_2){\otimes}\tilde{\bbox{e}}_A
                     (p_3){\otimes}\tilde{\bbox{e}}_A(p_4)
 \nonumber\\ 
   &+&\Gamma_{ARRR}(p_1,p_2,p_3,p_4) \,
   \tilde{\bbox{e}}_A(p_1){\otimes}\tilde{\bbox{e}}_R
                     (p_2){\otimes}\tilde{\bbox{e}}_R
                     (p_3){\otimes}\tilde{\bbox{e}}_R(p_4)
 \nonumber\\ 
   &+&\Gamma_{ARAA}(p_1,p_2,p_3,p_4)\,
   \tilde{\bbox{e}}_A(p_1){\otimes}\tilde{\bbox{e}}_R
                     (p_2){\otimes}\tilde{\bbox{e}}_A
                     (p_3){\otimes}\tilde{\bbox{e}}_A(p_4)
 \nonumber\\ 
   &+&\Gamma_{RARR}(p_1,p_2,p_3,p_4)\,
   \tilde{\bbox{e}}_R(p_1){\otimes}\tilde{\bbox{e}}_A
                     (p_2){\otimes}\tilde{\bbox{e}}_R
                     (p_3){\otimes}\tilde{\bbox{e}}_R(p_4)
 \nonumber\\
   &+&\Gamma_{AARA}(p_1,p_2,p_3,p_4)\,
   \tilde{\bbox{e}}_A(p_1){\otimes}\tilde{\bbox{e}}_A
                     (p_2){\otimes}\tilde{\bbox{e}}_R
                     (p_3){\otimes}\tilde{\bbox{e}}_A(p_4)
 \nonumber\\ 
   &+&\Gamma_{RRAR}(p_1,p_2,p_3,p_4)\,
   \tilde{\bbox{e}}_R(p_1){\otimes}\tilde{\bbox{e}}_R
                     (p_2){\otimes}\tilde{\bbox{e}}_A
                     (p_3){\otimes}\tilde{\bbox{e}}_R(p_4)
 \nonumber\\ 
   &+&\Gamma_{AAAR}(p_1,p_2,p_3,p_4)\,
   \tilde{\bbox{e}}_A(p_1){\otimes}\tilde{\bbox{e}}_A
                     (p_2){\otimes}\tilde{\bbox{e}}_A
                     (p_3){\otimes}\tilde{\bbox{e}}_R(p_4)
 \nonumber\\ 
   &+&\Gamma_{RRRA}(p_1,p_2,p_3,p_4)\,
   \tilde{\bbox{e}}_R(p_1){\otimes}\tilde{\bbox{e}}_R
                     (p_2){\otimes}\tilde{\bbox{e}}_R
                     (p_3){\otimes}\tilde{\bbox{e}}_A(p_4)
 \nonumber\\ 
   &+&\Gamma_{RRAA}(p_1,p_2,p_3,p_4)\,
   \tilde{\bbox{e}}_R(p_1){\otimes}\tilde{\bbox{e}}_R
                     (p_2){\otimes}\tilde{\bbox{e}}_A
                     (p_3){\otimes}\tilde{\bbox{e}}_A(p_4)
 \nonumber\\ 
   &+&\Gamma_{AARR}(p_1,p_2,p_3,p_4) \,
   \tilde{\bbox{e}}_A(p_1){\otimes}\tilde{\bbox{e}}_A
                     (p_2){\otimes}\tilde{\bbox{e}}_R
                     (p_3){\otimes}\tilde{\bbox{e}}_R(p_4)
 \nonumber\\ 
   &+&\Gamma_{ARRA}(p_1,p_2,p_3,p_4)\,
   \tilde{\bbox{e}}_A(p_1){\otimes}\tilde{\bbox{e}}_R
                     (p_2){\otimes}\tilde{\bbox{e}}_R
                     (p_3){\otimes}\tilde{\bbox{e}}_A(p_4)
 \nonumber\\ 
   &+&\Gamma_{RAAR}(p_1,p_2,p_3,p_4)\,
   \tilde{\bbox{e}}_R(p_1){\otimes}\tilde{\bbox{e}}_A
                     (p_2){\otimes}\tilde{\bbox{e}}_A
                     (p_3){\otimes}\tilde{\bbox{e}}_R(p_4)
 \nonumber\\
   &+&\Gamma_{ARAR}(p_1,p_2,p_3,p_4)\,
   \tilde{\bbox{e}}_A(p_1){\otimes}\tilde{\bbox{e}}_R
                     (p_2){\otimes}\tilde{\bbox{e}}_A
                     (p_3){\otimes}\tilde{\bbox{e}}_R(p_4)
 \nonumber\\ 
   &+&\Gamma_{RARA}(p_1,p_2,p_3,p_4)\,
   \tilde{\bbox{e}}_R(p_1){\otimes}\tilde{\bbox{e}}_A
                     (p_2){\otimes}\tilde{\bbox{e}}_R
                     (p_3){\otimes}\tilde{\bbox{e}}_A(p_4)\, .
 \label{Gamma4RA}
 \end{eqnarray}
The $R/A$ vertex functions satisfy the general definition
 \begin{equation}
 \label{gendefn}
   \Gamma_{\alpha_1\alpha_2\dots\alpha_n}(p_1,p_2,\dots,p_n) =
   \Bigl(\bbox{e}_{\alpha_1}(p_1){\otimes}\bbox{e}_{\alpha_2}
         (p_2){\otimes}\dots{\otimes}\bbox{e}_{\alpha_n}(p_n)\Bigr)
   \bullet \Gamma(p_1,p_2,\dots,p_n) 
 \end{equation}
(with $\alpha_i=R,A$) in terms of contractions with the basis vectors
(\ref{basis}). Eq.~(\ref{KMSgen}) gives the relations  
 \begin{mathletters}
 \label{KMS4}
 \begin{eqnarray}
 \label{KMS4a}
   \Gamma_{ARRR}(p_1,p_2,p_3,p_4) &=& 
   {(1+n_2)(1+n_3)(1+n_4)-n_2n_3n_4 \over (1+n_1) - n_1}
   \Gamma^*_{RAAA}(p_1,p_2,p_3,p_4)\, ,
 \\
 \label{KMS4b}
   \Gamma_{RARR}(p_1,p_2,p_3,p_4) &=& 
   {(1+n_1)(1+n_3)(1+n_4)-n_1n_3n_4 \over (1+n_2) - n_2}
   \Gamma^*_{ARAA}(p_1,p_2,p_3,p_4)\, ,
 \\
 \label{KMS4c}
   \Gamma_{RRAR}(p_1,p_2,p_3,p_4) &=& 
   {(1+n_1)(1+n_2)(1+n_4)-n_1n_2n_4 \over (1+n_3) - n_3}
   \Gamma^*_{AARA}(p_1,p_2,p_3,p_4)\, ,
 \\
 \label{KMS4d}
   \Gamma_{RRRA}(p_1,p_2,p_3,p_4) &=& 
   {(1+n_1)(1+n_2)(1+n_3)-n_1n_2n_3 \over (1+n_1) - n_1}
   \Gamma^*_{AAAR}(p_1,p_2,p_3,p_4)\, ,
 \\
 \label{KMS4e}
   \Gamma_{AARR}(p_1,p_2,p_3,p_4) &=& 
   {(1+n_3)(1+n_4)-n_3n_4 \over (1+n_1)(1+n_2) - n_1n_2}
   \Gamma^*_{RRAA}(p_1,p_2,p_3,p_4)\, ,
 \\
 \label{KMS4f}
   \Gamma_{ARRA}(p_1,p_2,p_3,p_4) &=& 
   {(1+n_2)(1+n_3)-n_2n_3 \over (1+n_1)(1+n_4) - n_1n_4}
   \Gamma^*_{RAAR}(p_1,p_2,p_3,p_4)\, ,
 \\
 \label{KMS4g}
   \Gamma_{ARAR}(p_1,p_2,p_3,p_4) &=& 
   {(1+n_2)(1+n_4)-n_2n_4 \over (1+n_1)(1+n_3) - n_1n_3}
   \Gamma^*_{RARA}(p_1,p_2,p_3,p_4)\, ,
 \end{eqnarray}
 \end{mathletters}
which reduce the number of independent retarded/advanced 4-point
functions to $2^{4-1}-1=7$. Combining (\ref{Gamma4RA}) with
(\ref{KMS4}) we get 
 \begin{eqnarray}
   &&\Gamma(p_1,p_2,p_3,p_4) =
 \nonumber\\ &&\phantom{+} 
   \Gamma_{RAAA}\,
   \tilde{\bbox{e}}_R(p_1){\otimes}\tilde{\bbox{e}}_A
                     (p_2){\otimes}\tilde{\bbox{e}}_A
                     (p_3){\otimes}\tilde{\bbox{e}}_A(p_4)
 \nonumber\\ 
   &&+{(1+n_2)(1+n_3)(1+n_4)-n_2n_3n_4 \over (1+n_1) - n_1}
   \Gamma^*_{RAAA}\,
   \tilde{\bbox{e}}_A(p_1){\otimes}\tilde{\bbox{e}}_R
                     (p_2){\otimes}\tilde{\bbox{e}}_R
                     (p_3){\otimes}\tilde{\bbox{e}}_R(p_4)
 \nonumber\\ 
   &&+\Gamma_{ARAA}\,
   \tilde{\bbox{e}}_A(p_1){\otimes}\tilde{\bbox{e}}_R
                     (p_2){\otimes}\tilde{\bbox{e}}_A
                     (p_3){\otimes}\tilde{\bbox{e}}_A(p_4)
 \nonumber\\ 
   &&+{(1+n_1)(1+n_3)(1+n_4)-n_1n_3n_4 \over (1+n_2) - n_2}
   \Gamma^*_{ARAA}\,
   \tilde{\bbox{e}}_R(p_1){\otimes}\tilde{\bbox{e}}_A
                     (p_2){\otimes}\tilde{\bbox{e}}_R
                     (p_3){\otimes}\tilde{\bbox{e}}_R(p_4)
 \nonumber\\
   &&+\Gamma_{AARA}\,
   \tilde{\bbox{e}}_A(p_1){\otimes}\tilde{\bbox{e}}_A
                     (p_2){\otimes}\tilde{\bbox{e}}_R
                     (p_3){\otimes}\tilde{\bbox{e}}_A(p_4)
 \nonumber\\ 
   &&+{(1+n_1)(1+n_2)(1+n_4)-n_1n_2n_4 \over (1+n_3) - n_3}
   \Gamma^*_{AARA}\,
   \tilde{\bbox{e}}_R(p_1){\otimes}\tilde{\bbox{e}}_R
                     (p_2){\otimes}\tilde{\bbox{e}}_A
                     (p_3){\otimes}\tilde{\bbox{e}}_R(p_4)
 \nonumber\\ 
   &&+\Gamma_{AAAR}\,
   \tilde{\bbox{e}}_A(p_1){\otimes}\tilde{\bbox{e}}_A
                     (p_2){\otimes}\tilde{\bbox{e}}_A
                     (p_3){\otimes}\tilde{\bbox{e}}_R(p_4)
 \nonumber\\ 
   &&+{(1+n_1)(1+n_2)(1+n_3)-n_1n_2n_3 \over (1+n_1) - n_1}
   \Gamma^*_{AAAR}\,
   \tilde{\bbox{e}}_R(p_1){\otimes}\tilde{\bbox{e}}_R
                     (p_2){\otimes}\tilde{\bbox{e}}_R
                     (p_3){\otimes}\tilde{\bbox{e}}_A(p_4)
 \nonumber\\ 
   &&+\Gamma_{RRAA}\,
   \tilde{\bbox{e}}_R(p_1){\otimes}\tilde{\bbox{e}}_R
                     (p_2){\otimes}\tilde{\bbox{e}}_A
                     (p_3){\otimes}\tilde{\bbox{e}}_A(p_4)
 \nonumber\\ 
   &&+{(1+n_3)(1+n_4)-n_3n_4 \over (1+n_1)(1+n_2) - n_1n_2}
   \Gamma^*_{RRAA} \,
   \tilde{\bbox{e}}_A(p_1){\otimes}\tilde{\bbox{e}}_A
                     (p_2){\otimes}\tilde{\bbox{e}}_R
                     (p_3){\otimes}\tilde{\bbox{e}}_R(p_4)
 \nonumber\\ 
   &&+\Gamma_{RAAR}\,
   \tilde{\bbox{e}}_R(p_1){\otimes}\tilde{\bbox{e}}_A
                     (p_2){\otimes}\tilde{\bbox{e}}_A
                     (p_3){\otimes}\tilde{\bbox{e}}_R(p_4)
 \nonumber\\ 
   &&+{(1+n_2)(1+n_3)-n_2n_3 \over (1+n_1)(1+n_4) - n_1n_4}
   \Gamma^*_{RAAR}
   \tilde{\bbox{e}}_A(p_1){\otimes}\tilde{\bbox{e}}_R
                     (p_2){\otimes}\tilde{\bbox{e}}_R
                     (p_3){\otimes}\tilde{\bbox{e}}_A(p_4)
 \nonumber\\
   &&+\Gamma_{RARA}(p_1,p_2,p_3,p_4)\,
   \tilde{\bbox{e}}_R(p_1){\otimes}\tilde{\bbox{e}}_A
                     (p_2){\otimes}\tilde{\bbox{e}}_R
                     (p_3){\otimes}\tilde{\bbox{e}}_A(p_4)
 \nonumber\\ 
   &&+{(1+n_2)(1+n_4)-n_2n_4 \over (1+n_1)(1+n_3) - n_1n_3}
   \Gamma^*_{RARA}\,
   \tilde{\bbox{e}}_A(p_1){\otimes}\tilde{\bbox{e}}_R
                     (p_2){\otimes}\tilde{\bbox{e}}_A
                     (p_3){\otimes}\tilde{\bbox{e}}_R(p_4)\, .
 \label{Gamma4}
 \end{eqnarray}
The generalization is obvious. We have chosen the $R/A$ vertices with
the smallest number of $R$ indices as independent $n$-point functions;
among them the functions with only a single $R$ index are the fully
retarded linear response functions \cite{WH98}. Eq.~(\ref{KMSgen})
permits, however, to select any other set of $2^{n-1}-1$ independent
$n$-point functions and to rewrite Eq.~(\ref{Gamma4}) and its
generalizations accordingly.

%%%%%%%%%%%%%%%%%%%%%%%%%%%%%%%%%%%%%%%%%%%%%%%%%%%%%%%%%%%%%%%%%%%%%%
\section{Conclusions}
\label{sec5}
%%%%%%%%%%%%%%%%%%%%%%%%%%%%%%%%%%%%%%%%%%%%%%%%%%%%%%%%%%%%%%%%%%%%%

We have given relations for arbitrary $n$ between the real-time
thermal $n$-point functions in the single-time representation and  
the retarded/advanced $n$-point vertex functions introduced by
Aurenche and Becherrawy \cite{Aur}. These relations were expressed in
a very compact way in terms of an orthogonal basis of two 2-component
column vectors and their dual partners. The advantage of the $R/A$
representation is that it diagonalizes the single particle propagator, 
shifting the thermal distribution functions from the propagators to
the vertices \cite{Aur}. In our column vector representation the
thermal distribution functions appear only inside the column vectors. 
Their orthogonality properties implement certain relations among the
thermal distribution functions which lead to immediate cancellations
between various different contributions \cite{u&m} to complex Feynman
diagrams; in other approaches to real-time finite temperature
field theory these calculations often occur at a much later stage of
the calculation. Thus our representation leads to appreciable
simplifications for the Feynman diagram calculus in the real-time
formulation of thermal field theory. 

Our column vector representation gives the thermal components of the
time-ordered $n$-point functions as linear combinations of $2^{n-1}-1$
independent retarded/advanced $n$-point functions and their complex
conjugates. Although we have here considered only bosonic $n$-point
functions, the generalization of our results to fermionic fields is
straightforward and involves only the simple substitution $n \to -n_f$
in the column vectors. The useful relation between thermal
distribution functions given in the Appendix has been presented 
for the general case, too. Similar relations as those given here for
the truncated $n$-point vertex functions hold for the connected
$n$-point Green functions; the only change is that in that case the
signs of all lower components in the column vectors $\tilde{\bbox{e}}$
must be reversed.

\acknowledgments
The authors gratefully acknowledge fruitful discussions with
P.~Henning and M.~Thoma and appreciate the financial support by 
the Deutsche Forschungsgemeinschaft (DFG), the Bundesministerium f\"ur
Bildung und Forschung (BMBF), the National Natural Science Foundation of
China (NSFC), and the Gesellschaft f\"ur Schwerionenforschung
(GSI). E.W. gratefully acknowledges support by the Alexander von
Humboldt Foundation through a Research Fellowship.

%%%%%%%%%%%%%%%%%%%%%%%%%%%%%%%%%%%%%%%%%%%%%%%%%%%%%%%%%%%%%%%%%%%
\appendix
%%%%%%%%%%%%%%%%%%%%%%%%%%%%%%%%%%%%%%%%%%%%%%%%%%%%%%%%%%%%%%%%%%%
\section{A useful relation}
\label{appa}
%%%%%%%%%%%%%%%%%%%%%%%%%%%%%%%%%%%%%%%%%%%%%%%%%%%%%%%%%%%%%%%%%%%

Let us write the thermal equilibrium distributions as
 \begin{equation}
 \label{A1}
    n(x_i) = {1\over e^{x_i} - \eta_i}\, ,\qquad
    n(-x_i) = -\eta_i - n(x_i)\, .
 \end{equation}
where $x_i = \beta p_i^0$, $\eta_i=1$ for bosons, and $\eta_i=-1$ for
fermions. Defining
 \begin{equation}
 \label{A2}
    \tilde{n}(\sum_{i=1}^k x_i) = 
    {1\over \exp\left(\sum_{i=1}^k x_i \right) -\prod_{i=1}^k \eta_i}
 \end{equation}
one can prove the identity
 \begin{equation}
 \label{A3}
   \prod_{i=1}^k \bigl(1+\eta_i n(x_i)\bigr) -
   \prod_{i=1}^k \eta_i n(x_i) =  
   {\prod_{i=1}^k n(x_i) \over \tilde{n}(\sum_{i=1}^k x_i)}\, .
 \end{equation}
It is easy to check the identity for $k=2$ and prove it for $k>2$ by
induction.

Due to fermion number conservation, an $n$-point function with $n$
external legs has always an even number of fermionic legs. This
implies 
 \begin{equation}
 \label{A4}
   \prod_{i=1}^n \eta_i = 1\, .
 \end{equation}
Since momentum conservation also requires 
 \begin{equation}
 \label{A5}
   \sum_{i=1}^n x_i = 0\, ,
 \end{equation}
Eq.~(\ref{A3}) yields the identity
 \begin{equation}
 \label{A6}
   \prod_{i=1}^n \bigl(1+\eta_i n(x_i)\bigr) -
   \prod_{i=1}^n  n(x_i) = 0 
 \end{equation}
if the product goes over all legs of an $n$-point function. Previously
used identities like Eq.~(34) in Ref.~\cite{u&m} are simple
consequences of Eq.~(\ref{A6}). The physical interpretation of
Eq.~(\ref{A6}) is that the probability for creating and for destroying
$n$ particles in the heat bath are equal. 

%%%%%%%%%%%%%%%%%%%%%%%%%%%%%%%%%%%%%%%%%%%%%%%%%%%%%%%%%%%%%%%%%%%
% 
% References:
%
%%%%%%%%%%%%%%%%%%%%%%%%%%%%%%%%%%%%%%%%%%%%%%%%%%%%%%%%%%%%%%%%%%%


\begin{references}
\bibitem{fn1}
  This can be seen by comparing the column vector representations of
  the propagator (\ref{Dp}) and self-energy (\ref{Pip}) and
  generalizing from the 2-point to the $n$-point case.
\end{references}
\end{document}